\def\tSNR{t_{\hbox{\sixrm SNR}}}
          \font\sixrm=cmr6       
               \font\sixbf=cmbx6      
\def\apj{{\it Ap. J. }}                                
\def\apjl{{\it Ap. J. Lett. }}                         
\def\aap{{\it Astron. Astr. }}                         
\def\app{{\it Astroparticle Phys. }}                   
\def\jetp{{\it Sov. Phys. JETP }}                      
\def\mnras{{\it MNRAS }}                               
\def\nat{{\it Nature }}                                
\def\ssr{{\it Space Sci. Rev. }}                       
\def\teq#1{$\, #1\,$}                         
\def\dover#1#2{\hbox{${{\displaystyle#1 \vphantom{(} }\over{
   \displaystyle #2 \vphantom{(} }}$}}                
{\catcode`\@=11                                                  
\gdef\SchlangeUnter#1#2{\lower2pt\vbox{\baselineskip 0pt\lineskip0pt    
\ialign{$\m@th#1\hfil##\hfil$\crcr#2\crcr\sim\crcr}}}}           
\def\gtrsim{\mathrel{\mathpalette\SchlangeUnter>}}

\documentstyle[psfig]{aipproc}

\begin{document}
\newcommand{\vol}[2]{$\,$\bf #1\rm , #2}                 
\vphantom{p}
\vskip -55pt
\centerline{\hfill To appear in Proc. of Snowbird TeV Gamma-Ray Workshop}
\centerline{\hfill ed. B.~L. Dingus (AIP, New York, 2000)}
\vskip 15pt

\title{Modelling Hard Gamma-Ray Emission\\ From Supernova Remnants}

\author{Matthew G.~Baring$^{\dagger}$}
\address{Laboratory for High Energy Astrophysics, Code 661\\
NASA Goddard Space Flight Center, Greenbelt, MD 20771\\
{\it baring@lheavx.gsfc.nasa.gov}\\
$^{\dagger}$Universities Space Research Association}

\maketitle

\begin{abstract}
The observation by the CANGAROO experiment of TeV emission from SN
1006, in conjunction with several instances of non-thermal X-ray
emission from supernova remnants, has led to inferences of super-TeV
electrons in these extended sources.  While this is sufficient to
propel the theoretical community in their modelling of particle
acceleration and associated radiation, the anticipated emergence in the
next decade of a number of new experiments probing the TeV and sub-TeV
bands provides further substantial motivation for modellers.  In
particular, the quest for obtaining unambiguous gamma-ray signatures of
cosmic ray ion acceleration defines a ``Holy Grail'' for observers and
theorists alike.  This review summarizes theoretical developments in
the prediction of MeV--TeV gamma-rays from supernova remnants over the
last five years, focusing on how global properties of models can
impact, and be impacted by, hard gamma-ray observational programs,
thereby probing the supernova remnant environment.  Properties of
central consideration include the maximum energy of accelerated
particles, the density of the unshocked interstellar medium, the
ambient magnetic field, and the relativistic electron-to-proton ratio.
Criteria for determining good candidate remnants for observability in
the TeV band are identified.
\end{abstract}

\section*{Introduction}
 \label{sec:intro}

It is widely believed that supernova remnants (SNRs) are the primary
sources of cosmic-ray ions and electrons up to energies of at least
\teq{\sim 10^{15}} eV, where the so-called {\it knee} in the spectrum
marks a deviation from almost pure power-law behavior.  Such cosmic
rays are presumed to be generated by diffusive (also called first-order
Fermi) acceleration at the remnants' forward shocks.  These cosmic rays
can generate gamma rays via interactions with the ambient interstellar
medium, including nuclear interactions between relativistic and cold
interstellar ions, by bremsstrahlung of energetic electrons colliding
with the ambient gas, and inverse Compton (IC) emission off cosmic
background radiation.  Rudimentary models of gamma-ray production in
remnants involving nuclear interactions date back to the late 1970s
\cite{hl75,chev77}.  These preceded the first tentative associations of
two COS-B gamma-ray sources \cite{poll85} with the remnants
\teq{\gamma} Cygni and W28.  Apart from the work of Dorfi
\cite{dorfi91}, who provided the first model including a more
sophisticated study of non-linear effects of shock acceleration to
treat gamma-ray production, the study of gamma-ray SNRs remained
quietly in the background until the observational program of the EGRET
experiment aboard the Compton Gamma Ray Observatory.  This provided a
large number of unidentified sources above 50 MeV, a handful of which
have interesting associations with relatively young SNRs
\cite{espos96}.

Following the EGRET advances, the modelling of gamma-ray and other
non-thermal emission from supernova remnants ``burgeoned,'' beginning
with the paper of Drury, Aharonian, \& V\"olk \cite{dav94} (hereafter
DAV94), who computed the photon spectra expected from the decay of
neutral pions generated in collisions of power-law shock-accelerated
ions with those of the interstellar medium (ISM).  This work spawned a
number of subsequent papers that used different approaches, as
discussed in the next section, and propelled the TeV gamma-ray
astronomy community into a significant observational program given the
prediction of substantial TeV fluxes from the DAV94 model.  The initial
expectations of TeV gamma-ray astronomers were dampened by the lack of
success of the Whipple and HEGRA groups \cite{less95,prosch96,buck97}
in detecting emission from SNRs after a concerted campaign.  While
sectors of the community contended that the constraining TeV upper
limits posed difficulties for SNR shock acceleration models, these
observational results were naturally explained
\cite{mdeJ96,baring97,sturn97} by the maximum particle energies
expected (in the 1--50 TeV range) in remnants and the concomitant
anti-correlation between maximum energy of gamma-ray emission and the
gamma-ray luminosity \cite{bergg99} (discussed below).

The observational breakthrough in this field came with the recent
report of a spatially-resolved detection of SN1006 (not accessible by
northern hemisphere atmospheric \v{C}erenkov telescopes (ACTs) such as
Whipple and HEGRA) by the CANGAROO experiment \cite{Tani98} at energies
above 1.7 TeV.  The interpretation (actually predicted for SN 1006 by
\cite{mdeJ96,Pohl96}) that evolved was that this emission was due to
energetic electrons accelerated in the low density environs of this
high-latitude remnant, generating flat-spectrum inverse Compton
radiation seeded by the cosmic microwave background.  This suggestion
was influenced, if not motivated by the earlier detection \cite{Koya95}
of the steep non-thermal X-ray emission from SN 1006 that has been
assumed to be the upper end of a broad synchrotron component, implying
the presence of electrons in the 20--100 TeV range.  Studies of
gamma-ray emission from remnants have adapted to this discovery by
suggesting (e.g.  \cite{baring97,bergg99}) that galactic plane remnants
such as Cas A that possess denser interstellar surroundings may have
acceleration and emission properties distinct from high-latitude
sources; the exploration of such a contention may be on the horizon,
given the detection of Cas A by HEGRA announced at this meeting
\cite{Voelk99}.  Given the complexity of recent shock acceleration/SNR
emission models, the range of spectral possibilities is considerable,
and a source of confusion for both theorists and observers.  It is the
aim of this paper to elucidate the study of gamma-ray remnants by
pinpointing the key spectral variations/trends with changes in model
parameters, and thereby identify the principal parameters that impact
TeV astronomy programs.

\section*{Models: A Brief History}
 \label{sec:history}

Reviews of recent models of gamma-ray emission from SNRs can be found
in \cite{baring97,bergg99,deJbar97,Voelk97}; a brief exposition is
given here.  Drury, Aharonian, \& V\"olk \cite{dav94} provided impetus
for recent developments when they calculated gamma-ray emission from
protons using the time-dependent, two-fluid analysis (thermal ions plus
cosmic rays) of \cite{dmv89}, following on from the similar work of
\cite{dorfi91}.  They assumed a power-law proton spectrum, so that no
self-consistent determination of spectral curvature to the
distributions \cite{eich84,ee84,je91} or temporal or spatial limits to
the maximum energy of acceleration was made.  The omission of
environmentally-determined high energy cutoffs in their model was a
critical driver for the interpretative discussion that ensued.
\cite{dav94} found that during much of Sedov evolution, maximal
diffusion length scales are considerably less than a remnant's shock
radius.

Gaisser, et al. \cite{gps98} computed emission from bremsstrahlung,
inverse Compton scattering, and pion-decay from proton interactions,
but did not consider non-linear shock dynamics or time-dependence and
assumed test-particle power-law distributions of protons and electrons
with arbitrary \teq{e/p} ratios.  In order to suppress the flat inverse
Compton component and thereby accommodate the EGRET observations of
$\gamma$ Cygni and IC443, \cite{gps98} obtained approximate constraints
on the ambient matter density and the primary \teq{e/p} ratio.

A time-dependent model of gamma-ray emission from SNRs using the Sedov
solution for the expansion was presented by Sturner, et al.
\cite{sturn97}. They numerically solved equations for electron and
proton distributions subject to cooling by inverse Compton scattering,
bremsstrahlung, \teq{\pi^0} decay, and synchrotron radiation (to supply
a radio flux).  Expansion dynamics and non-linear acceleration effects
were not treated, and power-law spectra were assumed.  Sturner et al.
(1997) introduced cutoffs in the distributions of the accelerated
particles (following \cite{mdeJ96,Reyn96,deJm97}), which are defined by
the limits (discussed below) on the achievable energies in Fermi
acceleration.  Hence, given suitable model parameters, they were able
to accommodate the constraints imposed by Whipple's upper limits
\cite{buck97} to $\gamma$ Cygni and IC 443.

To date, the two most complete models coupling the time-dependent
dynamics of the SNR to cosmic ray acceleration are those of Berezhko \&
V\"olk \cite{bv97}, based on the model of \cite{byk96}, and Baring et
al.  \cite{bergg99}.  Berezhko \& V\"olk numerically solve the gas
dynamic equations including the cosmic ray pressure and Alfv\'en wave
dissipation, following the evolution of a spherical remnant in a
homogeneous medium.  Originally only pion decay was considered, though
this has now been extended \cite{bkp99} to include other components.
Baring et al. simulate the diffusion of particles in the environs of
steady-state planar shocks via a well-documented Monte Carlo technique
\cite{je91,ebj96} that has had considerable success in modelling
particle acceleration at the Earth bow shock \cite{emp90} and
interplanetary shocks \cite{boef97} in the heliosphere.  They also
solve the gas dynamics numerically, and incorporate the principal
effects of time-dependence through constraints imposed by the Sedov
solution.

These two refined models possess a number of similarities.  Both
generate upward spectral curvature (predicted by \cite{eich84}; see the
review in \cite{baring97}), a signature that is a consequence of the
higher energy particles diffusing on larger scales and therefore
sampling larger effective compressions, and both obtain overall
compression ratios \teq{r} well above standard test-particle
Rankine-Hugoniot values.  Yet, there are two major differences between
these two approaches.  First Berezhko et al. \cite{bv97,bkp99} include
time-dependent details of energy dilution near the maximum particle
energy self-consistently, while Baring et al. \cite{bergg99} mimic this
property by using the Sedov solution to constrain parametrically the
maximum scale of diffusion (defining an escape energy).  These two
approaches merge in the Sedov phase \cite{eb99}, because particle
escape from strong shocks is a fundamental part of the non-linear
acceleration process and is determined primarily by energy and momentum
conservation, not time-dependence or a particular geometry.  Second,
\cite{bergg99} injects ions into the non-linear acceleration process
automatically from the thermal population, and so determine the
dynamical feedback self-consistently, whereas \cite{bv97} must specify
the injection efficiency as a free parameter.  Berezhko \& Ellison
\cite{eb99} recently demonstrated that, for most cases of interest, the
shock dynamics are relatively insensitive to the efficiency of
injection, and that there is good agreement between the two approaches
when the Monte Carlo simulation \cite{bergg99,ebj96} specifies injection
for the model of \cite{bv97}.  This convergence of results from two
complimentary methods is reassuring to astronomers, and underpins the
expected reliability of emission models to the point that a hybrid
``simple model'' has been developed \cite{be99} to describe the
essential acceleration features of both techniques.  This has been
extended to a new and comprehensive parameter survey \cite{ebb99} of
broad-band SNR emission that provides results that form the basis of
much of the discussion below.

\section*{Global Theoretical Predictions}
 \label{sec:predictions}

Since there is considerable agreement between the most developed
acceleration/emission models, we are in the comfortable position of
being able to identify the salient global properties that should be
characteristics of any particular model.  Clearly a treatment of
non-linear dynamics and associated spectral curvature are an essential
ingredient to more accurate predictions of emission fluxes,
particularly in the X-ray and gamma-ray bands where large dynamic
ranges in particle momenta are sampled, so that discrepancies of
factors of a few or more arise when test-particle power-laws are used.
Concomitantly, test-particle shock solutions considerably over-estimate
\cite{ebj96,be99,ebb99} the dissipational heating of the downstream
plasma in high Mach number shocks, thereby introducing errors that
propagate into predictions of X-ray emission and substantially
influence the overall normalization of hard X-ray to gamma-ray emission
(which depends on the plasma temperature \cite{bergg99,ebb99}).  These
points emphasize that a cohesive treatment of the entire particle
distributions is requisite for the accuracy of a given model.

In addition, finite maximum energies of cosmic rays imposed by
spatial and temporal acceleration constraints (e.g.
\cite{bergg99,berez96}) must be integral to any model, influencing
feedback that modifies the non-linear acceleration problem profoundly.
In SNR evolutionary scenarios, a natural scaling of this maximum energy
\teq{E_{\rm max}} arises, defined approximately by the energy attained
at the onset of the Sedov phase \cite{bergg99,berez96}:
\begin{equation}
   E_{\rm max} \;\sim\; 60\, \dover{Q}{\eta}\,
     \biggl(\dover{B_{\hbox{\sixrm ISM}}}{3\mu {\rm G}}\biggr)\,
     \biggl( \dover{n_{\hbox{\sixrm ISM}}}{1\, {\rm cm}^{-3}} \biggr)^{-1/3}\;
     \left( \dover{{\cal E}_{\hbox{\sixrm SN}}}{10^{51}{\rm erg}}\right)^{1/2}
     \,\biggl( \dover{M_{\rm ej}}{M_{\odot}} \biggr)^{-1/6} \ {\rm TeV}\;\; ,
 \label{eq:Emax}
\end{equation}
where \teq{Q} is the particle's charge, \teq{\eta} (\teq{\geq 1}) is
the ratio between its scattering mean-free-path and its gyroradius,
\teq{{\cal E}_{\hbox{\sixrm SN}}} is the supernova energy, \teq{M_{\rm
ej}} is its ejecta mass, and other quantities are self-explanatory.  At
earlier epochs, the maximum energy scales approximately linearly with
time, while in the Sedov phase, it slowly asymptotes
\cite{bergg99,bv99} to a value a factor of a few above that in
Eq.~(\ref{eq:Emax}).

\begin{figure}[t]
\centerline{\psfig{figure=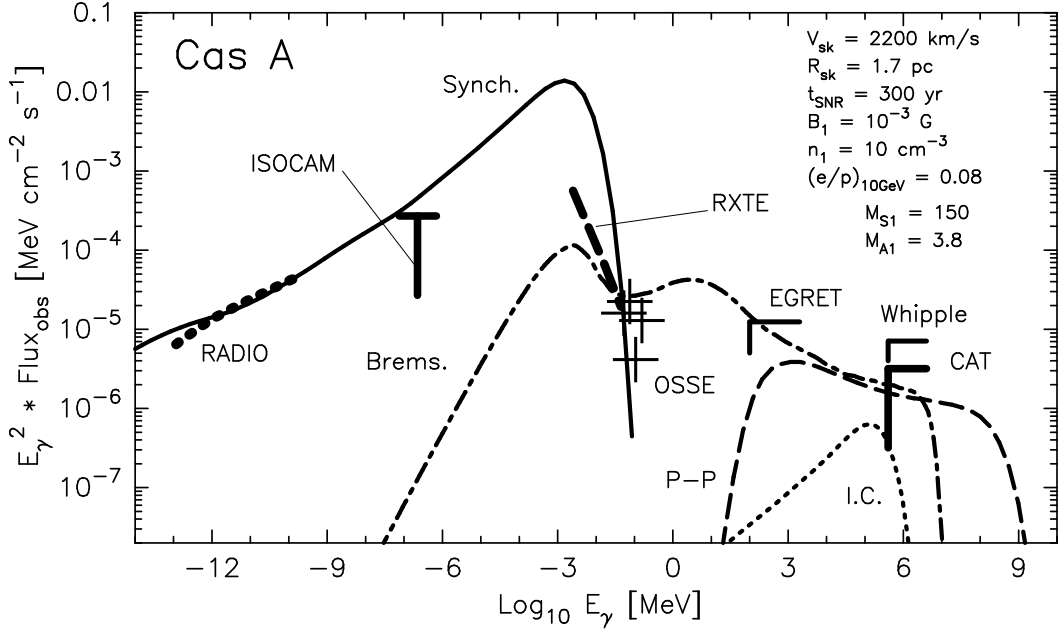,width=14.0truecm}}
\caption{The Cassiopeia A spectrum from the Monte Carlo acceleration
calculation of Ellison et al. ([38], see this for detailed
referencing of the data sources). The model photons come from a single
set of proton, helium, and electron spectra calculated with the
upstream parameters shown in the figure.  A single normalization factor
has been applied to all components to match the radio flux.  Note that
the bremsstrahlung and inverse Compton (IC) emission cuts off at a much
lower energy than the pion decay radiation due to the synchrotron
losses the electrons experience.  In these results, the IC component
does not include a synchrotron self-Compton contribution.
 \label{fig:CasAspec}}
\end{figure}

Three properties emerge as global signatures of models that impact
observational programs.  The first is that there is a strong
anti-correlation of \teq{E_{\rm max}} (and therefore the maximum energy
of gamma-ray emission) with gamma-ray luminosity, first highlighted by
\cite{bergg99}.  High ISM densities are conducive to brighter sources
in the EGRET to sub-TeV band \cite{bergg99,bv97,ebb99}, but reduce
\teq{E_{\rm max}} in Eq.~(\ref{eq:Emax}) and accordingly act to inhibit
detection by ACTs.  Low ISM magnetic fields produce a similar trend,
raising the gamma-ray flux by flattening the cosmic ray distribution
(discussed below).  Clearly, high density, low \teq{B_{\hbox{\sixrm
ISM}}} remnants are the best candidates for producing cosmic rays up to
the knee.  Fig.~\ref{fig:CasAspec} displays a sample model spectrum for
Cas A, which has a high density, high \teq{B_{\hbox{\sixrm ISM}}}
environment.  In it the various spectral components are evident, and
the lower \teq{E_{\rm max}} for electrons (relative to that for
protons) that is generated by strong cooling is evident in the
bremsstrahlung and inverse Compton spectra.

The other two global properties are of a temporal nature.  The first is
the approximate constancy of the observed gamma-ray flux (and
\teq{E_{\rm max}} \cite{bv99}) in time during Sedov phase, an
insensitivity first predicted by \cite{dorfi91} and confirmed in the
analyses of \cite{dav94,bergg99,bv99}.  The origin of this
insensitivity to SNR age \teq{\tSNR} is an approximate compensation
between the SNR volume \teq{{\cal V}} that scales as \teq{\tSNR^{6/5}}
(radius \teq{\propto\tSNR^{2/5}}) in the Sedov phase, and the
normalization coefficient \teq{{\cal N}} of the roughly \teq{E^{-2}}
particle distribution function: since the shock speed (and therefore
also the square root of the temperature \teq{T_{\rm pd}}) declines as
\teq{\tSNR^{-3/5}}, it follows that \teq{{\cal N}\propto T_{\rm
pd}\propto \tSNR^{-6/5}} and flux\teq{\propto {\cal N}{\cal
V}\approx}const.  There is also a limb brightening with age
\cite{dav94} that follows from the constant maximum particle length
scale concurrent with continuing expansion.

\subsection*{Key Parameters and Model Behavioural Trends}
 \label{sec:behaviour}

The principal aim here is to distill the complexity of non-linear
acceleration models for time-dependent SNR expansions and discern the
key parameters controlling spectral behaviour and simple reasons for
behavioural trends.  This should elucidate for theorist and
experimentalist alike the scientific gains to be made by present and
next generation experimental programs.  Parameters are grouped
according to them being of model origin and environmental nature
(trends associated with the age of a remnant were discussed just above),
and details can be found in the comprehensive survey of Ellison,
Berezhko \& Baring \cite{ebb99}.

There are three relevant {\it model} parameters in non-linear
acceleration, the ratio of downstream electron and proton temperatures
\teq{T_{\rm ed}/T_{\rm pd}}, the injection efficiency \teq{\eta_{\rm
inj}} (after \cite{bv97,byk96}), and the electron-to-proton ratio
\teq{(e/p)_{\rm rel}} at relativistic energies (i.e. \teq{\gtrsim
1}--\teq{10}GeV).  The injection efficiency is the most crucial of
these, since it controls the pressure contained in non-thermal ions,
and therefore the non-linearity of the acceleration process.  It mainly
impacts the X-ray to soft gamma-ray bremsstrahlung contribution, a
component that is generally dominated by pion decay emission in the
hard gamma-ray band.  The shape and normalization of the \teq{\pi^0}
decay gamma-rays is only affected when \teq{\eta_{\rm inj}} drops below
\teq{10^{-4}} and the shock solution becomes close to the test-particle
one, i.e. an overall spectral steepening arises.  Variations in
\teq{(e/p)_{\rm rel}} influence the strength of the inverse Compton and
bremsstrahlung components, which modify the total gamma-ray flux only
if \teq{(e/p)_{\rm rel}\gtrsim 0.1}, a high value relative to cosmic
ray abundances, or the ambient field is strong.

\begin{figure}
\centerline{\psfig{figure=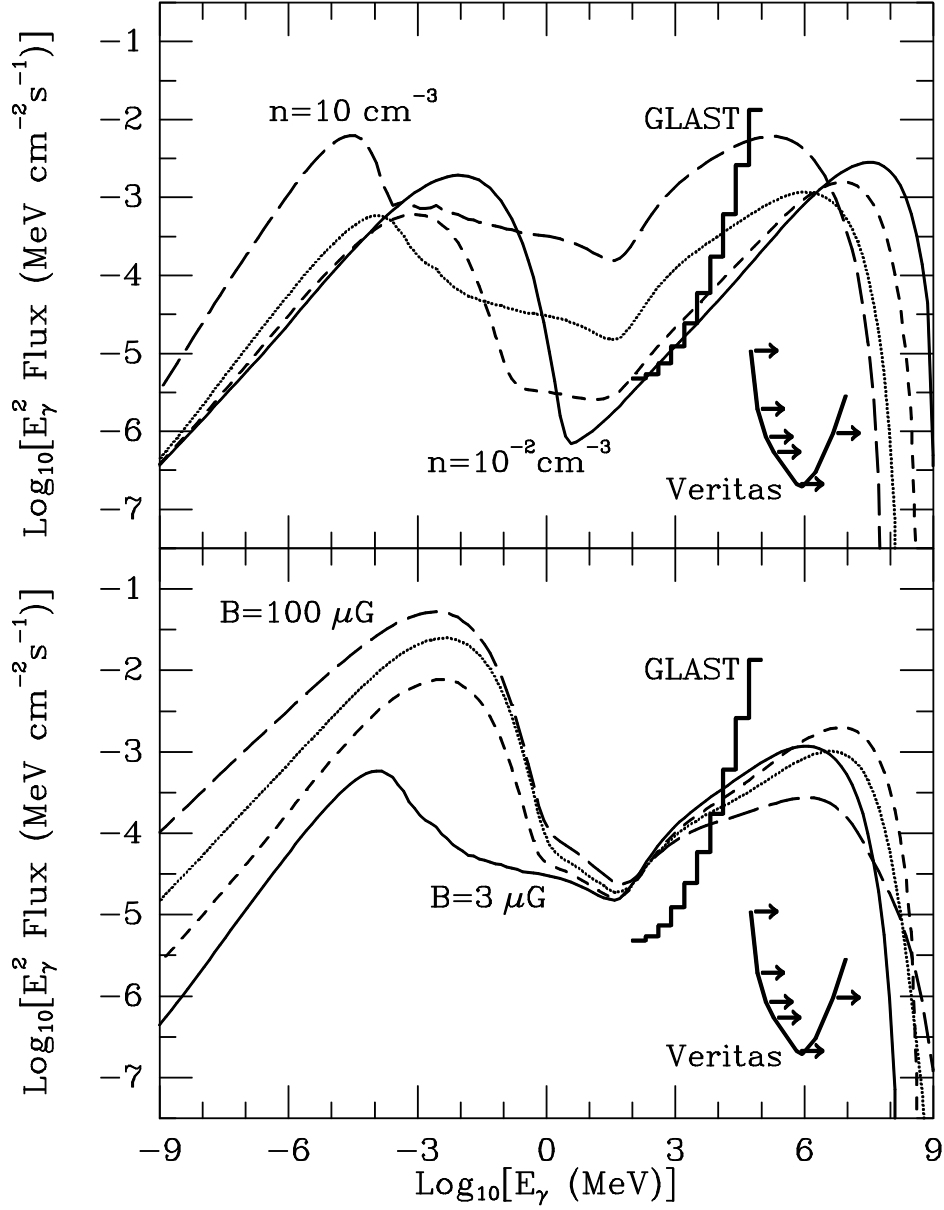,height=16.0truecm}}
\caption{Trends of total photon emission for variations of ISM
parameters \teq{n\equiv n_{\hbox{\sixrm ISM}}} and \teq{B\equiv
B_{\hbox{\sixrm ISM}}}, adapted from the simplified approximate
description of non-linear acceleration in [35].  Top panel: the ISM
field is fixed at $B=3\mu$G, and the ambient number density is varied
such that:  $n=0.01$ cm$^{-3}$ (solid), $n=0.1$ cm$^{-3}$ (short
dashes), $n=1$ cm$^{-3}$ (small dots), and $n=10$ cm$^{-3}$ (long
dashes).  Bottom panel: $B$ is varied:  $B=3\mu$G (solid), $B=10\mu$G
(short dashes), $B=30\mu$G (small dots), and $B=100\mu$G (long dashes),
with the density pinned to $n=1$ cm$^{-3}$.  Here \teq{(e/p)_{\rm
rel}=0.03}; consult [35] for other model parameters.  Also depicted are
the canonical integral flux sensitivity for Veritas [42]
and the differential flux sensitivity for GLAST (Digel,
private communication) to facilitate the discussion in the text.
  \label{fig:trends}}
\end{figure}

The most interesting behavioural trends are elicited by the {\it
environmental} parameters \teq{n_{\hbox{\sixrm ISM}}} and
\teq{B_{\hbox{\sixrm ISM}}}, and the results adapted from \cite{ebb99}
are illustrated in Fig.~\ref{fig:trends}.  Naively, one expects that
the radio-to-X-ray synchrotron and gamma-ray inverse Compton components
should scale linearly with density increases, while the bremsstrahlung
and pion decay contributions intuitively should be proportional to
\teq{n_{\hbox{\sixrm ISM}}^2}.  However, global spectral properties are
complicated by the non-linear acceleration mechanism and the evolution
of the SNR.  As \teq{n_{\hbox{\sixrm ISM}}} rises, the expanding
supernova sweeps up its ejecta mass sooner, and therefore decelerates
on shorter timescales, thereby reducing both the volume \teq{{\cal V}}
of a remnant of given age, and lowering the shock speed and the
associated downstream ion temperature \teq{T_{\rm pd}}.  Hence, the
density increase is partially offset by the ``shifting'' of the
particle distributions to lower energies (due to lower \teq{T_{\rm
pd}}) so that the normalization \teq{{\cal N}} of the non-thermal
distributions at a given energy is a weakly increasing function of
\teq{n_{\hbox{\sixrm ISM}}}.  Clearly \teq{{\cal V}} times this
normalization controls the observed flux of the synchrotron and inverse
Compton components, while the product of \teq{{\cal N}}, the target
density \teq{n_{\hbox{\sixrm ISM}}} and \teq{{\cal V}} determines the
bremsstrahlung and \teq{\pi^0\to\gamma\gamma} emission, with results
shown in Fig.~\ref{fig:trends}.  Observe that
the approximate constancy of the inverse Compton contribution
effectively provides a lower bound to the gamma-ray flux in the 1
GeV--1 TeV band, a property that is of significant import in defining
experimental goals.

The principal property in Fig.~\ref{fig:trends} pertaining to
variations in \teq{B_{\hbox{\sixrm ISM}}} is the anti-correlation
between radio and TeV fluxes: the higher the value of
\teq{B_{\hbox{\sixrm ISM}}}, the brighter the radio synchrotron, but
the fainter the hard gamma-ray pion emission.  This property is
dictated largely by the influence of the field on the shock dynamics
and total compression ratio \teq{r}: the higher the value of
\teq{B_{\hbox{\sixrm ISM}}}, the more the field contributes to the
overall pressure, reducing the Alfv\'enic Mach number and accordingly
\teq{r}, as the flow becomes less compressible.  This weakening of the
shock steepens the particle distributions and the overall photon
spectrum.  An immediate offshoot of this behaviour is the premise
\cite{ebb99} that radio-selected SNRs may not provide the best targets
for TeV observational programs.  Case in point: Cas A is a very bright
radio source while SN 1006 is not, and the latter was observed first.

Since \teq{n_{\hbox{\sixrm ISM}}} and \teq{B_{\hbox{\sixrm ISM}}}
principally determine the gamma-ray spectral shape, flux normalization
and whether or not the gamma-ray signatures indicating the presence of
cosmic ray ions are apparent, they are the most salient parameters to
current and future ACT programs and the GLAST experiment.

\section*{Key Issues and Experimental Potential}
 \label{sec:experiment}

There are a handful of quickly-identifiable key issues that define
goals for future experiments, and these can be broken down into two
categories: spatial and spectral.  First and foremost, the astronomy
community needs to know whether the EGRET band gamma-ray emission is
actually shell-related.  While the associations of \cite{espos96} were
enticing, subsequent research \cite{braz96,keohane97,braz98} has
suggested that perhaps compact objects like pulsars and plerions or
concentrated regions of dense molecular material may be responsible for
the EGRET unidentified sources.  If a connection to the shell is
eventually established, it is desirable to know if it is localized only
to portions of the shell.  One naturally expects that shock obliquity
effects \cite{ebj96} can play an important role in determining the
gamma-ray flux in ``clean'' systems like SN 1006, and that turbulent
substructure within the remnant (e.g. Cas A) can complicate the picture
dramatically.  Such clumping issues impact radio/gamma-ray flux
determinations, since the radio-emitting electrons diffuse on shorter
length scales and therefore are more prone to trapping.  Another
contention that needs observational verification is whether or not
limb-brightening increases with SNR age?  Improvements in the angular
resolution of ACTs can resolve these issues and discern variations in
gamma-ray luminosities across SNR shells: the typical capability of
planned experiments such as HESS, Veritas, MAGIC and CANGAROO-III is of
the order of 2--3 arcminutes in the TeV band \cite{weekes99,kohnle99}.

The principal gamma-ray spectral issue is whether or not there is
evidence of cosmic ray {\it ions} near remnant shocks.  The goal in
answering this is obviously the detection of \teq{\sim 70}MeV $\pi^0$
bump, the unambiguous signature of cosmic ray ions, and given the GLAST
{\it differential} sensitivity (the measure of capability in performing
spectral diagnostics as opposed to detection above a given energy)
plotted in Fig.~\ref{fig:trends}, GLAST will be sensitive to remnants
with \teq{n_{\hbox{\sixrm ISM}}\gtrsim 0.1} cm$^{-3}$.  Atmospheric
\v{C}erenkov experiments can also make progress on this issue, with the
dominant component in the super-TeV band for moderately or highly
magnetized remnant environs being that of pion decay emission (see
Fig.~\ref{fig:CasAspec}).  Such a circumstance may already be realized
in the recent marginal detection \cite{Voelk99} of Cas A by HEGRA.  The
most powerful diagnostic the sensitive TeV experiments will provide is
the determination of the maximum energy (see Fig.~\ref{fig:trends}) of
emission (and therefore also that of cosmic ray ions or electrons),
thereby constraining \teq{n_{\hbox{\sixrm ISM}}}, \teq{B_{\hbox{\sixrm
ISM}}} and the \teq{e/p} ratio.  Furthermore, the next generation of
ACTs should be able to discern the expected anti-correlation between
\teq{E_{\rm max}} and $\gamma$-ray flux, and with the help of GLAST,
search for spectral concavity, a principal signature of non-linear
acceleration theory.  In view of the anticipated increase in the number
of TeV SNRs, a population classification may be possible, determining
whether or not SN1006 and other out-of-the-plane remnants differ
intrinsically in their gamma-ray and cosmic ray production from the Cas
A-type SNRs.  These potential probes augur well for exciting times in
the next 5--10 years in the field of TeV gamma-ray astronomy.


\vskip 5pt\noindent
{\bf Acknowledgments:}  I thank my collaborators Don Ellison, Steve
Reynolds, Isabelle Grenier, Frank Jones and Philippe Goret for many
insightful discussions, Seth Digel for providing results of simulations
of GLAST spectral capabilities, and Rod Lessard for supplying the
Veritas integral flux sensitivity data for Fig.~\ref{fig:trends}.

\end{document}